\def\year{2022}\relax
\documentclass[letterpaper]{article} 
\usepackage{aaai22}  
\usepackage{times}  
\usepackage{helvet}  
\usepackage{courier}  
\usepackage[hyphens]{url}  
\usepackage{graphicx} 
\usepackage{natbib}  
\usepackage{caption} 
\DeclareCaptionStyle{rule
d}{labelfont=normalfont,labelsep=colon,strut=off} 
\usepackage{algorithm}
\usepackage{algorithmic}

\usepackage{newfloat}
\usepackage{listings}
\floatstyle{ruled}
\newfloat{listing}{tb}{lst}{}
\floatname{listing}{Listing}

\usepackage{amsmath}
\usepackage{xcolor}
\usepackage{comment}
\usepackage{subfig}
\title{Antibody Representation Learning for Drug Discovery}

\author{Lin Li \textsuperscript{\rm 1}, 
Esther Gupta \textsuperscript{\rm 1},
John Spaeth \textsuperscript{\rm 1},
Leslie Shing \textsuperscript{\rm 1},
Tristan Bepler \textsuperscript{\rm 2, 3},
Rajmonda Sulo Caceres \textsuperscript{\rm 1}}
\affiliations{
    \textsuperscript{\rm 1}MIT Lincoln Laboratory, 
    Lexington, MA 02421\\
    \textsuperscript{\rm 2}Research Laboratory of Electronics, MIT, Cambridge, 02139 \\
    \textsuperscript{\rm 3}Simons Machine Learning Center, NYSBC, New York, 10027 \\
    \{lin.li, esther.wolf, john.spaeth, leslie.shing\}@ll.mit.edu, {tbepler}@mit.edu, {rajmonda.caceres}@ll.mit.edu
}

\usepackage{bibentry}

\begin{document}

\maketitle

\renewcommand{\thefootnote}{\small}
\footnotetext{DISTRIBUTION STATEMENT A. Approved for public release. Distribution is unlimited. This material is based upon work supported by the Under Secretary of Defense for Research and Engineering under Air Force Contract No. FA8702-15-D-0001. Any opinions, findings, conclusions or recommendations expressed in this material are those of the author(s) and do not necessarily reflect the views of the Under Secretary of Defense for Research and Engineering. © 2021 Massachusetts Institute of Technology.
Delivered to the U.S. Government with Unlimited Rights, as defined in DFARS Part 252.227-7013 or 7014 (Feb 2014). Notwithstanding any copyright notice, U.S. Government rights in this work are defined by DFARS 252.227-7013 or DFARS 252.227-7014 as detailed above. Use of this work other than as specifically authorized by the U.S. Government may violate any copyrights that exist in this work.}

\begin{abstract}
Therapeutic antibody development has become an increasingly popular approach for drug development. To date, antibody therapeutics are largely developed using large scale experimental screens of antibody libraries containing hundreds of millions of antibody sequences. The high cost and difficulty of developing therapeutic antibodies create a pressing need for computational methods to predict antibody properties and create bespoke designs. However, the relationship between antibody sequence and activity is a complex physical process and traditional iterative design approaches rely on large scale assays and random mutagenesis. Deep learning methods have emerged as a promising way to learn antibody property predictors, but predicting antibody properties and target-specific activities depends critically on the choice of antibody representations and data linking sequences to properties is often limited. Recently, methods for learning biological sequence representations from large, general purpose protein datasets have demonstrated impressive ability to capture structural and functional properties of proteins. However, existing works have not yet investigated the value, limitations and opportunities of these methods in application to antibody-based drug discovery. In this paper, we present results on a novel SARS-CoV-2 antibody binding dataset and an additional benchmark dataset. We compare three classes of models: conventional statistical sequence models, supervised learning on each dataset independently, and fine-tuning an antibody specific pre-trained embedding model. The pre-trained models were trained on tens of millions of natural healthy human antibody sequences and protein sequences, respectively. Experimental results suggest that self-supervised pretraining of feature representation consistently offers significant improvement in over previous approaches. We also investigate the impact of data size on the model performance, and discuss challenges and opportunities that the machine learning community can address to advance \textit{in silico} engineering and design of therapeutic antibodies.
\end{abstract}

\noindent Drug development is a complex, time-consuming and costly process. The current drug design paradigm is primarily an assay driven process: millions or billions of candidate molecules are screened for activity to identify a small number of candidates for further development and optimization. Often, this involves intensive laboratory experiments and expert-guided analysis to design and test molecule variants. This approach poses high costs in terms of the equipment, experimental environment, and the expert knowledge required. For antibody drug development, candidate antibodies are typically identified by affinity maturation of enormous naive antibody libraries \textit{in vitro} with phage or yeast display or \textit{in vivo} by animal immunization~\cite{lu2020development}. These methods require target molecules (epitopes) to be stably producible and rely on sheer scale to identify antibody sequences that interact with the target. This makes these processes expensive and unpredictable, because there is no guarantee that a suitable antibody exists within the starting library and candidates can fail later in the drug development process due to safety or manufacturability concerns. As a result, the ability to identify many candidate molecules and optimize these quickly is critical for predictable, fast, and inexpensive drug development.

With several technological advances in experimental antibody development, including the ability to obtain pure antibodies in large numbers for research and clinical use, as well as, the successful translation of antibodies to the clinic~\cite{lu2020development}, antibody design--a predominant therapeutic modality for various diseases~\cite{lu2020development}--has become an increasingly popular approach for drug development. While these advancements have shortened timelines throughout the development pipeline, \textit{in silico} engineering of antibody candidates that enables cheaper and faster drug development against rapidly-evolving antigens still remains a challenge.
Computational methods for antibody development, to date, have largely focused on physical simulation-based or simple statistical model-based approaches to optimize the complementary determining regions (CDRs) of antibodies~\cite{tiller2015advances}. Physics-based methods approach antibody optimization from first principles by simulating interactions between antibodies and targets at the atomic level. Although these methods are conceptually appealing and offer the promise of generalization to any antibody or target sequence, in practice, they require solved structures for antibody-target complexes and are impractically slow to compute~\cite{yamashita2018toward}, limiting the space of antibody variants that can be computationally explored.

Machine learning (ML) has demonstrated potential in accelerating drug discovery with good performance and lower computational cost~\cite{yang2019analyzing,bepler2019learning,liu2020antibody,stokes2020deep}. A key challenge in ML for drug development is learning effective feature representations that capture structural and functional properties of candidate drug designs. Traditionally, ML models used in bioinformatics research rely on expert-designed descriptors to predict drug-like properties. These descriptors are often limited to known chemical or biochemical classes and are inherently challenging for unexplored systems. Recent advances of neural networks provide an alternative way to learn representations automatically from data. 
The work in~\cite{yang2019analyzing} has leveraged graph neural networks (GNNs) to learn representations of molecular data, which are found to offer significant improvements over models currently used in industrial workflows. Using a similar approach, authors in~\cite{stokes2020deep} have identified Halicin, a new antibiotic molecule that is effective against a broad class of disease-causing bacteria. 
Authors in~\cite{bepler2019learning,rao2019evaluating,alley2019unified,rives2021biological,bepler2021learning} have explored learning protein sequence representations in a self-supervised and  semi-supervised fashion to predict structural properties and other down-stream tasks. They have highlighted the value of various neural network architectures and pre-training in substantially improving our ability to predict structural and functional properties from sequences alone.
Authors in~\cite{liu2020antibody} have shown promising results in applying CNN architectures for both the task of antibody enrichment prediction and antibody design. While above works suggest promise of deep representation learning for biological systems, many questions still remain open for antibody-specific feature representation for drug discovery. 

 

In this work, we develop the first antibody-specific pre-trained protein embedding models and incorporate the learned features into a broader ML system for antibody binding affinity prediction. Using this system, we compare the descriptive and predictive power of conventional antibody features and features learned by deep language models. Next, we explore the role of pre-training in supporting knowledge transfer and improving antibody binding prediction performance, and we evaluate the importance of pre-training datasets by comparing models trained only on antibody sequences to models trained on general protein sequences. Finally, we investigate training data sufficiency requirements that support performance robustness and knowledge transfer. We analyze the sensitivity of feature learning as a function of training data size and characterize data size ranges that are most suitable for deep learning of antibody features.
We make the following contributions:

\begin{itemize}
\item We present two large antibody sequence datasets with binding affinity measurements~\citep{matthew_walsh_2021_5095284, liu2020antibody} for analyzing antibody-specific representation learning. 
The LL-SARS-CoV-2 dataset, in particular, is the first of its kind in terms of sequence diversity, consisting of SARS-Cov2 antibody variants generated from mutations across the full antibody sequence. 
\item We develop the first antibody sequence-specific pre-trained language model. We find that pre-training on general protein sequence datasets supports better feature refinement and learning for antibody binding prediction. This result highlights the important role of training over a diverse set of protein sequences.
\item We demonstrate that language models consistently learn much more effective features for antibody binding prediction than conventional antibody sequence features or features learned by a CNN model. This result is consistent with recent results that have investigated the power of language models in supporting downstream tasks in the natural language domain.~\cite{Saunshi-ICLR2021}.
\item Our training data size sensitivity analysis reveals a range of data sizes (about 3200-6400 samples), where we observe the most steep improvement of binding prediction performance. Fewer data samples are insufficient to capture the antibody sequence to binding mapping, while more data samples offer diminishing returns. 
\end{itemize}

\section{Background}
\begin{figure}[t!]
    \centering
        \includegraphics[width=1\linewidth]{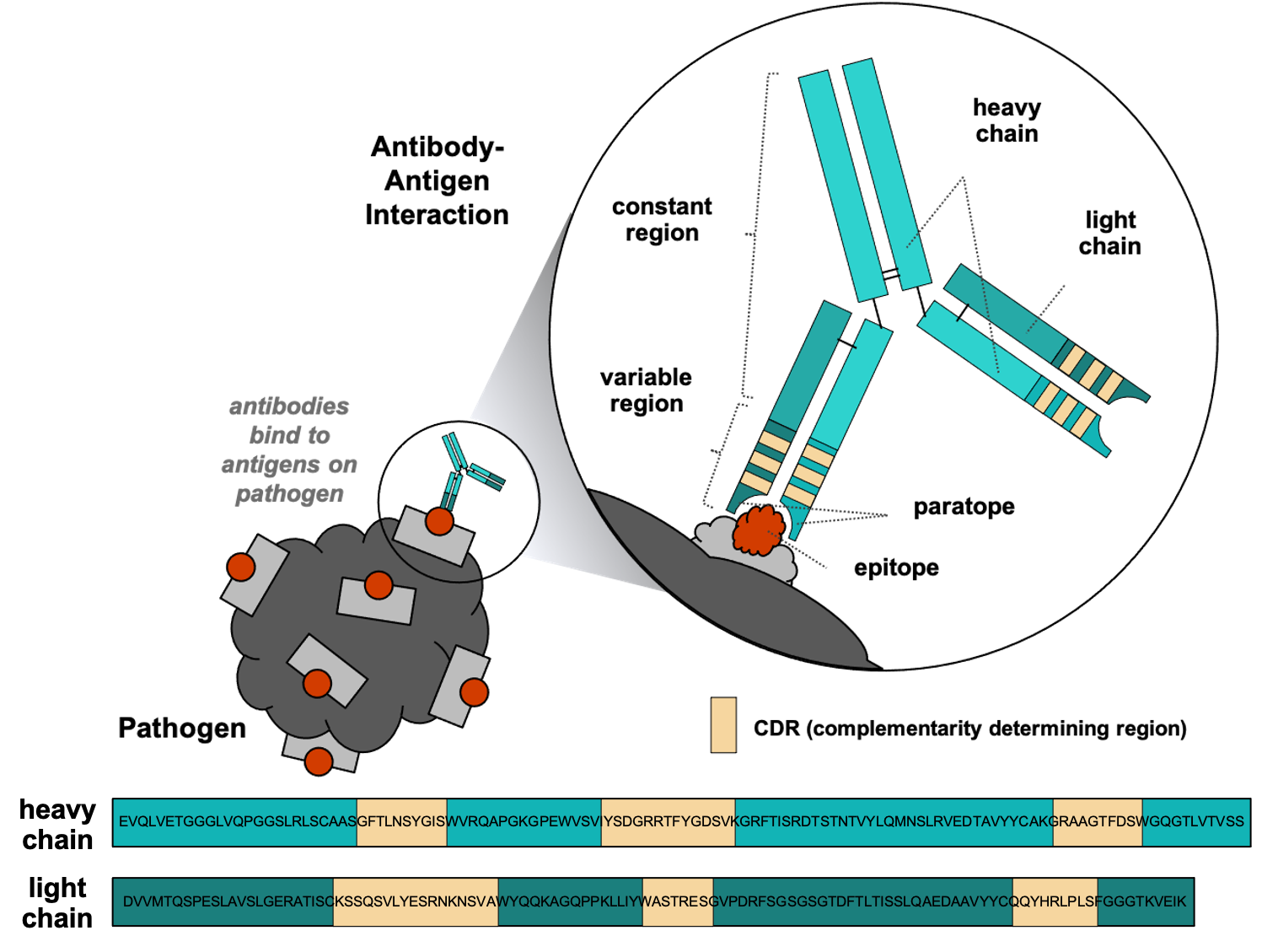}
    \caption{Illustration of the antibody structure and its interaction with an antigen.}
    \label{fig:antibody}
\end{figure}

Antibodies are Y-shaped proteins produced by the immune system to tag or neutralize foreign substances, called antigens.
As shown in Figure~\ref{fig:antibody}, antibodies consist of two identical light chains and two identical heavy chains. Each chain has a variable region and a constant region. The tip of the variable region forms the antibody's binding surface, also known as a paratope. This region recognizes and binds to a specific antigen's binding surface called an epitope. The variable regions of each chain contain three hypervariable 
regions known as the complementarity-determining regions (CDRs), denoted as CDR-L1, CDR-L2, CDR-L3 and CDR-H1, CDR-H2, CDR-H3, for the light and heavy chains respectively (highlighted in yellow in Figure~\ref{fig:antibody}). 

The amino acid sequence of the CDRs determines the antigens to which an antibody will bind~\cite{o2010essentials}. The variable domain of the light chain can consist of 94-139 amino acids, while that of the heavy chain is slightly longer, consisting of 92-160 amino acids~\cite{abhinandan2008analysis}. Each amino acid is encoded into a 25-character alphabet with $20$ characters for the standard amino acids and $5$ characters for non-standard, ambiguous or unknown amino acids. 

Many natural properties of antibodies are considered when designing optimized antibody candidates against a specific target including binding affinity, binding specificity, stability, solubility, and effector functions. 
Antibody binding affinity, the property we focus on in this paper, is the binding strength between the antibody and antigen, and is inversely related to the equilibrium dissociation constant $K_D$, a ratio of $K_{\rm off}$/$K_{\rm on}$, where $K_{\rm off}$ measures how quickly an antibody dissociates from the antigen target, and $K_{on}$ measures how quickly the antibody binds to the target~\cite{landry2015measuring}. 

Studies on protein crystal structures, including antibodies, have been integral in understanding protein-protein interactions, protein function, and therapeutic development. An understanding of how sequence relates to the structure of the variable domains may illuminate the effects on binding properties of antibody-antigen interactions. Approaches exist that attempt to improve antibody modeling by leveraging sequence data to determine the structure of the CDR regions, since the amount of available sequence data far exceeds that of structural data. These efforts include homology modeling for regions outside CDR-H3. Another modeling effort groups and classifies non-CDR-H3 regions into structural clusters by matching them to Position Specific Scoring Matrices (PSSMs) derived from multiple sequence alignments of known structural groups \cite{graves2020review}. Sequence alignment methods can help infer homologous relationships (e.g., common ancestry, structures and functions) between two or more sequences and underpin most computational protein analysis methods. 



\section{Antibody Datasets}
The continued development of advanced sequencing technologies in the past decade has led to exponential growth in the number of protein and antibody sequences~\cite{marks2020repertoire}. The wealth of sequence information provides a great opportunity for learning antibody features via self-supervised language models, which have achieved significant success in natural language processing (NLP) applications. The hypothesis is that statistical patterns in sequences contain important information about their associated functions and structures, can be captured in the learned representations of language models. 

In this section, we describe the datasets used to train antibody feature representation models. To evaluate and compare the learned representations and conventional antibody features, we have also curated several benchmark datasets for antibody binding prediction tasks. Each of these tasks differs in terms of binding targets, parts of the antibody sequences that are known, and ways of measuring the binding strength.  

\subsection{Unlabeled Sequence Data}
\subsubsection{Pfam.}

Pfam is a well-known database of curated protein families containing raw sequences of amino acids for individual protein domains \cite{pfam}. We train a general protein language model on Pfam family sequences following \cite{bepler2019learning,rao2019evaluating} using the splits provided by \cite{rao2019evaluating}. The train, validation, and test splits contain 32,593,668, 1,715,454, and 44311 sequences, respectively. 

\subsubsection{OAS database} 
The Observed Antibody Space (OAS) database is a project that gathers and annotates immune repertoires for use in large scale analyses \cite{kovaltsuk2018observed}. It contains a compilation of over one billion raw antibody sequences from over 75 studies and is routinely updated. The full database contains immune repertoires from a diverse set of immune states (i.e. specific diseases or na{\"i}ve/disease-free) and provides both unpaired (i.e. individual light and heavy chains) and a small number of paired antibody sequences. 

We curated a subset of these sequences to perform self-supervised language model pre-training to learn the antibody-specific feature space representations. The antibody sequence search space is a subset of the full protein sequence search space, so this dataset provides learned representations at a more granular level.  At the time of download, we included only studies with na{\"i}ve human subjects and randomly removed redundant sequences across the studies for individual light and heavy chains. De-duplication efforts resulted in 9 studies containing 70,838,791 light chain sequences and 37 studies containing 270,171,931 heavy chain sequences. The train/validation/test  sequence splits were determined based on the studies (i.e. sequences from the same study were kept in the same split). Specifically, the light chain data splits contain 70,059,824 train, 364,332 validation and 414,635 test sequences and the heavy chain data splits contain 172,524,747 train, 46,603,347 validation and 51,043,837 test sequences.

\subsection{Labeled Sequence Data}
ML models rely heavily on the size and quality of the training data. However, there remains a scarcity of labeled antibody data available in the public domain. The lack of labeled data is largely due to the data requirements when conducting biological experiments. For example, phage display \cite{hoogenboom1998antibody}, a technique for high-throughput screening of protein interactions, is intended to find a small number of top binders from a large pool of sequences. It does not provide quantitative measurements that are needed for training ML models. While recent methods, such as AlphaSeq \cite{younger2017high} that use engineered yeast expression systems and advanced DNA sequencing, can generate large scale datasets with quantitative binding measurements, most of these proprietary datasets generally only exist in the private domain or are generated as part of high throughout experimental campaigns. 

In this section, we discuss two large-scale labeled datasets that contain antibody sequences and associated binding affinity measurements. The first is a dataset that contains quantitative binding scores of antibodies against a SARS-CoV-2 target peptide using an AlphaSeq assay~\cite{matthew_walsh_2021_5095284}, and the second dataset contains affinity enrichment measurements for a library of CDR-H3 sequences against the Ranibizumab antibody fragment target~\cite{liu2020antibody}. 

\subsubsection{LL-SARS-CoV-2 data}
We utilize 
a labeled binding affinity dataset of antibody sequences against a SARS-CoV-2 target, a conserved peptide in coronavirus \cite{matthew_walsh_2021_5095284}. The affinity measurement is represented as the $log10$ of the dissociation constant $K_D$; smaller values indicate higher binding affinities. 
Specifically, the dataset was generated using a high-throughput approach for \textit{in silico} antibody library design and antibody binding affinity measurements. Antibodies were derived from three antibody backbones, namely Ab-14, Ab-91, and Ab-95, determined through phage display and initial affinity measurements. The authors selected two heavy chains and two light chains from the three antibody backbones to serve as seeds for the \textit{in silico} design of the antibody libraries: 14H, 91H, 14L, and 95L, respectively, where `H' denotes the heavy chain and `L' denotes the light chain of the selected antibody backbone. Each of the four chains includes all three CDR regions, along with the static framework regions between the CDRs. The \textit{in silico} design process generated $29,900$ sequence variants for each chain, introducing $K$-point mutations (i.e. substitutions), where $K=1,2$ and $3$, across all CDR regions for each variant, generating a total of $119,600$ designs. Finally, the AlphaSeq technique was applied to each sample variant to provide affinity binding measurements for each sequence variant against the SARS-CoV-2 target. Each sample resulted in three replicate binding affinity measurements. Average of three replicate measurements is used as the affinity label. 
This resulted in four datasets, one for each of the variants of 14H, 14L, 91H and 95L. The full number of unique samples in each dataset are 25,474 for 14L, 25,210 for 14H, 21,577 for 91H, and 29,384 for 95L. The datasets differ in peptide length, but within each dataset the peptide length is equal. The samples were randomly split into train/validation/test sets with $0.8$/$0.1$/$0.1$ split. 


\subsubsection{IgG Antibody dataset.}
We also utilize a dataset of CDR-H3 region sequences and labeled with affinity measurements against the ranibizumab target, a monoclonal antibody that is aimed at slowing the growth of abnormal new blood vessels. The authors of \cite{liu2020antibody} chose to use antibodies as the antigen targets in their work because the known common and unique regions of the target antibodies aided in the goals of their ML efforts. The authors developed a phage display library, diversifying only the CDR-H3 region, which varied in length between 10-16 and 18 amino acids and performed phage display panning using that library against the target. Three rounds of panning and sequencing occurred, resulting in the expected reduction of sequence diversity in each round as washing procedures intensified to reduce the noise and identify candidate sequences based on phage enrichment. Only the CDR-H3 regions were segmented out. The procedure resulted in 572,647 unique CDR-H3 sequences against the target in round 1, 297,290 in round 2, and and 171,568 in round 3. The final regression dataset uses the round 2 and round 3 panning enrichment measurements as a label, and is a subset of the whole pool of sequences. The train, validation, and test sizes are 60,992, 6,777, and 29,007 respectively. The peptides are variable length, with lengths ranging from 8 to 20.



\section{Feature Representations}
In this section, we summarize several classes of non-ML and ML-learned feature representations for antibodies and how they are used in training ML models for prediction. 

\subsubsection{Conventional Feature Representations}
Position-specific scoring matrix (PSSM) is one of the most commonly used representations in biological sequences. Given a set of $N$ functionally related sequences in a multiple sequence alignment (fixed sequence length $L$), i.e., $S_n = [x_i, x_2, \cdots, x_L]$ and $n \in (1, \cdots, N)$,  
\begin{align}
PSSM =   \begin{bmatrix}
   p_{1,1}& \cdots & p_{1,20}\\
      \vdots & \ddots & \vdots \\
   p_{L,1} & \cdots & p_{L,20}
   \end{bmatrix}
\end{align}
where the column represents the $20$ standard amino acid vocabulary $[a_1, a_2, \cdots, a_{20}]$, and $p_{i,j}$ denotes the probability of the amino acid in the $i^{th}$ position of the sequence mutates into the amino acid $a_j$ in the process of biological evolution. Specifically, let $M$ be a position probability matrix estimated by counting  amino acid frequencies at each of the sequence position. Then $p_{i,j} = \log_2 \frac{M(i,j)}{b_j}$ and $b_j$ represents a background model. Assuming each amino acid letter appears equally frequently in the dataset, then we have $b_j = 1/20$.

The next step is to extract a sequence-specific vector representation from PSSM, we follow the approach proposed in~\cite{zahiri2013ppievo}.  That is, each sequence is represented by a vector $400$ elements 
$$V = [v_1^{(1)}, \cdots, v_{20}^{(1)},v_1^{(2)}, \cdots, v_{20}^{(2)}, \cdots, v_1^{(20)}, \cdots, v_{20}^{(20)}]$$
where $v_j^{(i)}$ denotes the sum of all substitute score from $a_i$ to $a_j$, i.e., $v_j^{(i)} = \sum_{\ell=1}^L p_{\ell, j} * \delta_{\ell, i}$ with $\delta =1$ if $x_\ell = a_i$ and zero otherwise. The PSSM-based features are fixed in length and can be directly used as inputs to a ML model. 

\subsubsection{Supervised Representation Learning}
Advances in neural networks allow features to be learned automatically from labeled data and use them to perform learning tasks. With sufficient number of labeled data, convolutional neural networks (CNN) and Long Short-Term Memory (LSTM) have been used to train antibody feature representation for affinity prediction~\cite{liu2020antibody,saka2021antibody}. In both models, one-hot encoding vectors are used as inputs. The features extracted from the model are inputs to the decision head which is usually a densely connected feed-forward network layer. While both models have shown promising results when applied to antibody learning tasks, in this paper, we choose to use CNN as the representative model for supervised representation learning due to its successful application in antibody CDR design~\cite{liu2020antibody}. 

\subsubsection{Self-Supervised Language Models}
With the vast amount of unlabeled biological sequences, self-supervised learning has received significant attention in recent years to learn sequence representations~\cite{rao2019evaluating}. In particular, multiple self-supervised language models have been trained, evaluated and compared on a set of protein prediction benchmarks, ranging from structure prediction and evolutionary understanding to protein engineering. The pretrained language models include BERTTransformer and dilated residual network (ResNet) that are trained with masked-token prediction~\cite{vaswani2017attention,yu2017dilated}, and LSTM-based models that are trained with next-token prediction~\cite{hochreiter1997long,bepler2019learning,alley2019unified}. The learned features are embedding vectors for each amino acid in the sequence. To obtain a sequence-length invariant embedding, an attention-weighted mean of amino-acid embeddings is used as the input to the final decision layer in the fine-tuning step. In the case where the downstream ML model cannot be combined with neural networks to perform fine-tuning, one can first freeze amino-acid embeddings and then compute the mean of these embeddings. Alternatively, if the sequences are of the same length, one can concatenate the amino acid embeddings, followed by dimensionality reduction.

The performance of self-supervised language models is mostly determined by the amount of training data, the quality of the data and the choice of the language model. The work in~\cite{rao2019evaluating} has performed an extensive analysis on comparing language models on a variety of protein learning tasks, and results show that for protein engineering tasks, the BERTTransformer masked language model outperforms other model architectures considerably. However, there is still a need to understand how training data effects the language model performance, particularly for antibodies. To address this, we investigate the BERTTransformer language model trained on three different datasets: Pfam dataset, OAS light chain sequences and OAS heavy chain sequences. While Pfam covers a large collection of protein families, including antibodies, the OAS data are more specific to antibodies. The goal is to understand the impact of data granularity on pre-trained language models and how it effects the model transferrability to downstream tasks.

\subsection{Training Size Dependence}
Apart ML techniques for antibody representation learning, various other factors contribute to the applicability of ML in drug discovery. Most importantly, the performance of ML models relies heavily on the size of the training data. However, in practice, we are often limited by the number of labeled data that are publicly available. For example, existing antibody sequence-to-property mappings range from a few hundreds~\cite{Yoon2015} to a hundred thousand sequences~\citep{matthew_walsh_2021_5095284,liu2020antibody}. In contrast, more traditional applications of language models leverage much larger labeled training sets, from millions to billions of data samples~\cite{devlin-bert2018}. Even in these settings, recent research~\cite{Tamkin2020,kaplan2020} has highlighted scaling laws and limitations of language models trained with insufficient data.
Therefore, it is important to understand the effects of the training size on ML performance and on the robustness of various antibody feature representations, a question we investigate in this paper.

\section{Experiments}\label{sec:exp}
\subsection{Experimental Setup}
\subsubsection{Model Architecutres}

The BERTTransformer is used to train the self-supervised language model. The input embedding size is $768$, and there are $24$ hidden layers, with a hidden size of $1024$ and intermediate feed-forward size of $4096$. The number of attention heads is $16$. For the parameters not listed, we use the default values that were used in~\cite{rao2019evaluating}. 
During pre-training, the learning rate is set to $1^{-5}$, the batch size is $1024$ and the warm-up step is $10000$. 
To use features learned from the pre-trained model for antibody binding prediction, we analyzed two approaches: (1) fine-tune the BERTTransformer model with a standard multi-layer perceptron (MLP) decision layer; (2) freeze the embeddings and train a ML model (e.g., ridge repression and Gaussian process regression). For model fine-tuning, the hidden dimension of the MLP layer is set at $512$, half of the input feature dimensions, and the drop-out rate is set to 0.1. 

The CNN is used to train the supervised representation model. It is structured as close as possible to the largest CNN used in~\cite{liu2020antibody}. Each input token is converted into a one-hot vector, where each position of the vector is a channel input to the CNN. The entire dataset is zero padded to the largest sequence length of the dataset. The model is composed of two convolutional layers followed by a standard MLP decision layer. Each convolutional layer consists of a convolution operation, a max pool operation, and a ReLU activation. Both convolution operations have a kernel size of 5 and stride of 1. The convolution operations have 32 and 64 filters, respectively. Both max pool operations have a kernel size of 2 and stride of 2. The dense layer has 16 neurons with ReLU activation with a .15 dropout rate. The hidden dimension of the MLP decision layer is set to be half of the input dimension.

All models are train on NVIDIA Volta V100 GPUs using a distributed compute architecture. 

\subsubsection{Evaluation Metrics}
The BERTTransformer language model is a masked-token prediction task, which models the probability of the masked token given the unmasked tokens. We use the standard average perplexity to evaluate the language model performance on the test data. Note that the perplexity only measures how well the language model predicts the masked token, and it does not necessarily correspond to better performance on downstream tasks. 

To compare the performance of antibody feature representations on downstream tasks and investigate sensitivities of these models to training data, we evaluate these representations against a set of benchmark datasets on antibody binding prediction tasks. Pearson correlation is used as the performance metric.

\subsubsection{Datasets}
All models trained for the affinity binding prediction task are evaluated on three distinct antibody datasets: 14H and 14L datasets from LL-SARS-Cov-2, and IgG antibody dataset. Note that while 14H and 14L datasets consist of heavy chain and light chain variants of the Ab-14 backbone, respectively, the IgG includes only the CDR-H3 region of an antibody backbone. 

\subsection{Pre-trained Model Comparison}
To understand how training data effects the performance of language models, we trained three BERTTransformer models. They are (1) the Pfam model trained on the Pfam dataset; (2) the heavy chain model trained on the OAS heavy chain sequences; and (3) the light chain model trained on the OAS light chain sequences. The language models are first evaluated on their respective test data. The average perplexities are $13.1508$, $1.5990$ and $1.4316$ for the Pfam model, heavy chain model and light chain model, respectively. 

To compare these pre-trained language models on the downstream task, we explore two transfer learning approaches. One is to fine-tune the pre-trained model on the downstream regression task. For this approach, an attention-weighted mean of amino-acid embeddings is used followed by a standard multilayer percetron (MLP) decision layer, then we train the entire network initialized with the weights from the pre-trained model. The second approach is to freeze the amino-acid embedding layer and then construct a sequence-length invariant embedding. Specifically, for 14H and 14L, since antibody sequences are aligned and of the same length, we concatenated the embeddings of amino acids in all three CDR regions and performed dimensionality reduction with principal component analysis (PCA). For IgG, since the sequences have variable length, we simply computed the mean of all amino acid embeddings. 

Table~\ref{tab:featrep_finetune} shows the Pearson correlation results on the three antibody binding affinity prediction datasets via fine-tuning the pre-trained model. Table~\ref{tab:featrep_gp} shows the affinity prediction performance when training a Gaussian process (GP) model on features extracted from the pre-pretrained model. Gaussian process regression is especially relevant in drug design because it allows us to quantify prediction uncertainty which help focus the experimental effort on drug candidates with high likelihood of success.

\begin{table}[th!]
\centering
\scalebox{0.9}{
    \begin{tabular}{|l|l|l|l|}
    \hline
     &{\bf Heavy Chain}& {\bf Light Chain}  & {\bf Pfam}\\ \hline
    14H & 0.646 & 0.642 &  {\bf 0.664}\\ \hline
    14L &0.662&0.670 & {\bf 0.692} \\ \hline
     IgG (CDR-H3)& {\bf 0.835} & 0.832 & {\bf 0.835} \\ 
    \hline
    \end{tabular}}
    \caption{Pearson correlations on predicting antibody binding affinities via fine-tuning pre-trained language models}
    \label{tab:featrep_finetune} 
\end{table}

\begin{table}[th!]
\centering
\scalebox{0.9}{
    \begin{tabular}{|l|l|l|l|}
    \hline
     &{\bf Heavy Chain}& {\bf Light Chain}  & {\bf Pfam}\\ \hline
    14H & 0.588 & 0.602 &  {\bf 0.638}\\ \hline
    14L &0.657&0.612 & {\bf 0.678} \\ \hline
     IgG (CDR-H3)& 0.733 & 0.710 & {\bf 0.757} \\ 
    \hline
    \end{tabular}}
    \caption{Pearson correlations on predicting antibody binding affinities via training a Gaussian process model over features extracted from pre-trained language models}
    \label{tab:featrep_gp} 
\end{table}

We observe that across different prediction tasks, the language model trained on the more diverse Pfam data consistently outperforms the models trained on antibody-specific light and heavy chain data. This indicates the importance of diversity in training data. While protein sequences may exhibit differing underlying distributions, knowledge gained from learning the protein language model captures higher-level biological principles that can be transferred and refined to antibody-specific tasks. For the rest of the experiments on feature representation comparison, the pre-trained language model refers to the model trained on the Pfam data. 

\subsection{Comparison of Feature Representation Models}
\begin{figure}[t]%

    \centering
    \subfloat[Multilayer Perceptron (MLP) Decision Head]{{\includegraphics[scale=.43]{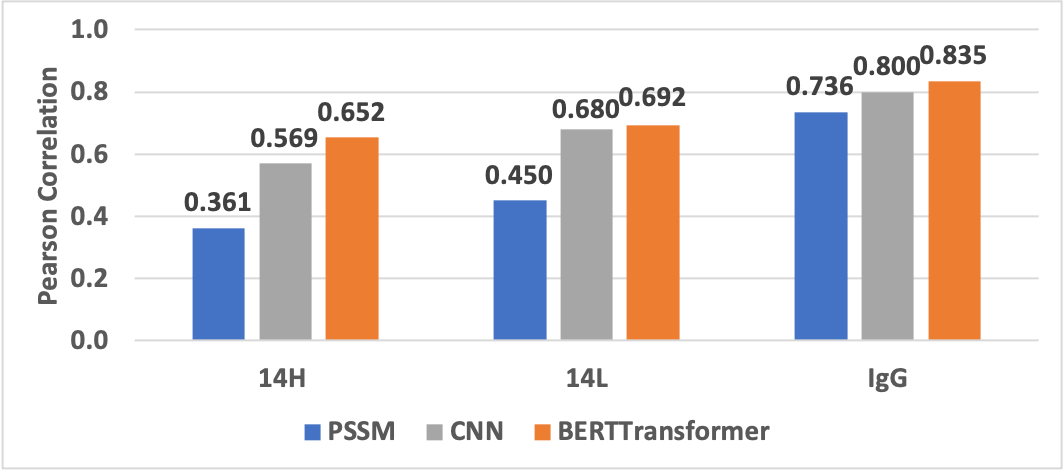} }}%
    
    \subfloat[Ridge Regression]{{\includegraphics[scale=.43]{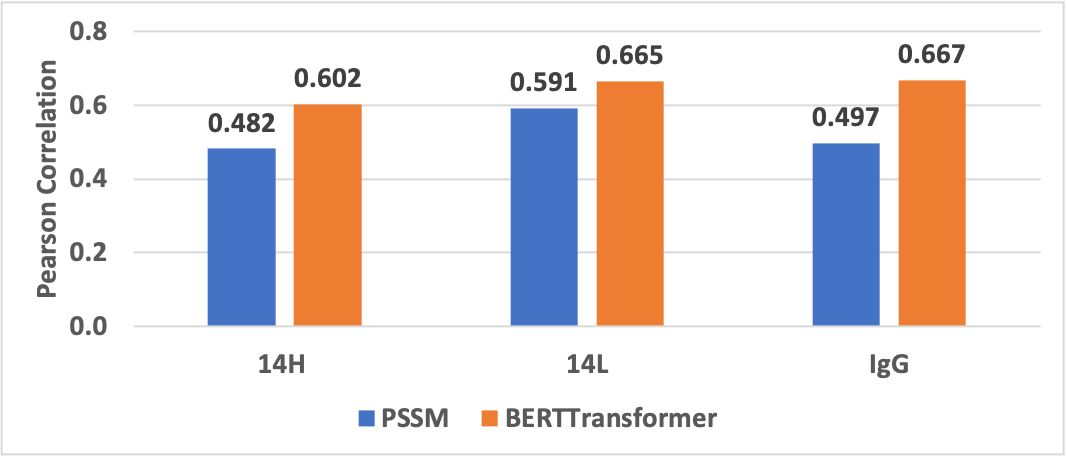} }}%
    
    \subfloat[Gaussian Process (GP) Regression
    ]{{\includegraphics[scale=.43]{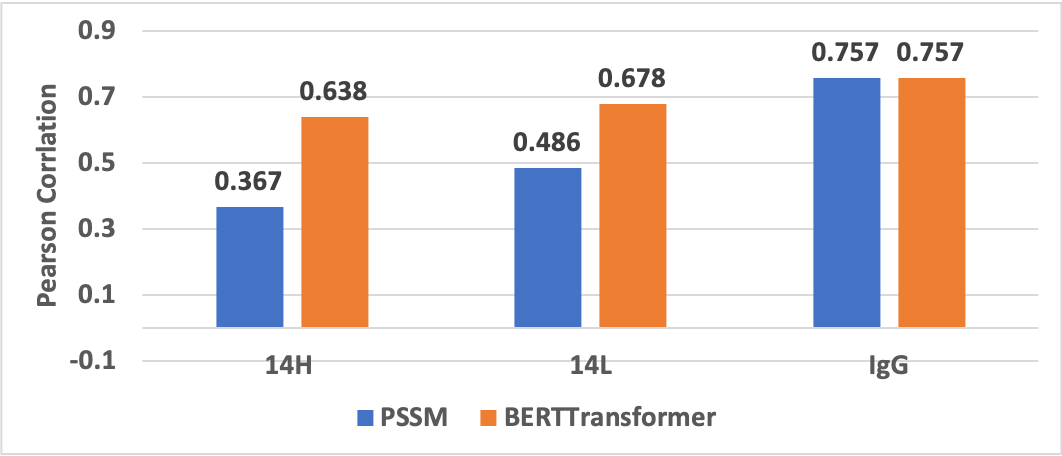} }}%
    
    \caption{Feature representation comparison for 14H, 14L and IgG affinity prediction tasks}%
    \label{fig:comparison_feat}%
    
\end{figure}
We compare three classes of feature representations: conventional PSSM-based features, features produced by supervised learning on the labeled data, and features learned by the pre-trained language model.  
To compare these features, we explore different regression models: MLP, ridge regression and Gaussian process (GP) regression.  Figure~\ref{fig:comparison_feat} shows the prediction performance, respectively. Overall, the results across all learning experiments indicate that  the pre-pretrained model consistently learns better antibody features for binding predictions than other feature learning approaches. The performance of PSSM-based features is highly dependent on the regression models and training tasks. For example, the ridged regression model performs better on 14L and 14H tasks while Gaussian process performs better on IgG task using the PSSM-based features. 

\subsection{Training Data Size Dependence}
One crucial factor when comparing model performance, is the amount of training data required. Not only does the size of the dataset dictate which ML techniques are successful but many organizations must appropriately budget their investment in data collection. We investigate the dependency of the training data size on the model performance of the antibody binding affinity prediction task.

The model performance is obtained from five independent runs for each training size using a different train/validation sub-samples and averaging the Pearson correlations. For each run, the train and validation data are randomly sub-sampled, where the validation sub-sample is $10 \%$ of the training sub-sample.  The same random train/validation splits are used to train all models compared. Hyperparameter tuning on the learning rate is also performed on each run using $10$ trials of random grid search. The search space for the learning rate is $10^{-4}$ to $10^{-6}$ for finetuning pretrained models and $10^{-2}$ to $10^{-4}$ for CNN and GP models. All models are evaluated on the full test set. 

\begin{figure}[htbp]%

    \centering
    \subfloat[14L]{{\includegraphics[scale=.29]{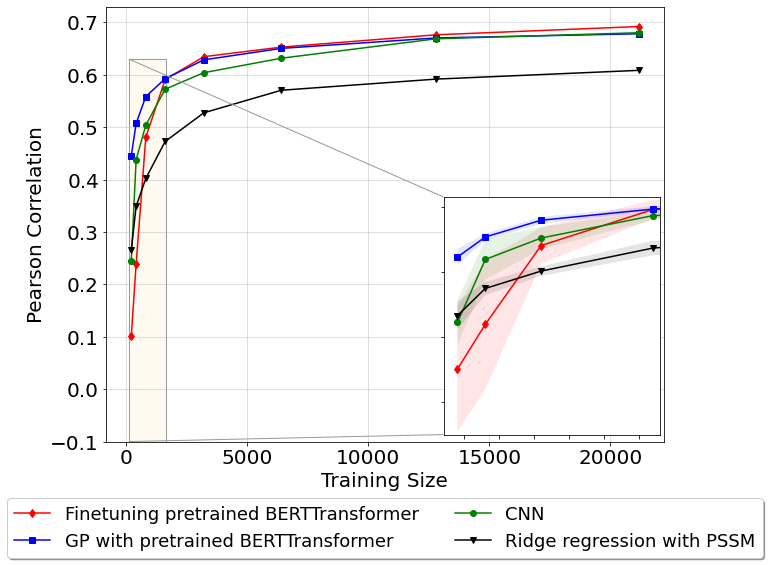} }}%
    
    \subfloat[14H]{{\includegraphics[scale=.29]{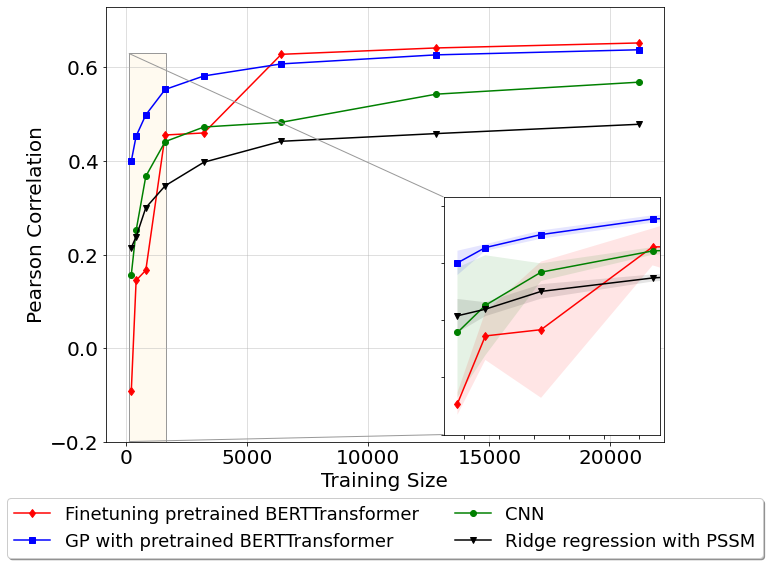} }}%
    
    \subfloat[IgG]{{\includegraphics[scale=.29]{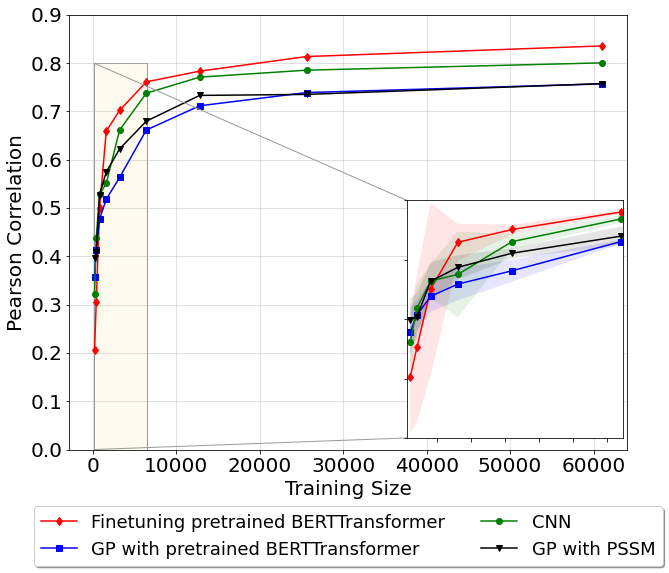} }}%
    
    \caption{The training size dependence for 14L, 14H and IgG are plotted. Each point represents the average of five separate runs using random train/validation sub-samples. The shaded area in the zoomed-in plot represents the standard deviation. The training data sizes in the zoomed-in plots are 200-1600 for 14L and 14H, and 200-6400 for IgG.}%
    \label{fig:titration}%
    
\end{figure}

Figure~\ref{fig:titration} shows the model performance on 14L, 14H, and  with IgG datasets. For 14L and 14H, the ridge regression is trained on the PSSM-based features, while for IgG, Gaussian process (GP) regression is used as GP performs much better  on the PSSM-based features than other regression models; see Figure~\ref{fig:comparison_feat}. Overall, when the training data is sufficient, fine-tuning the pre-trained language consistently outperforms other models. For a small set of training data, GP trained over features extracted from the pre-trained model tends to perform better with less performance variance over different training sub-samples, and it also outperforms the the conventional method, indicating the robustness of the pre-trained model in support of transfer learning. Additionally, training sizes between 3200 and 6400 provide the most steep performance improvement, which can be used to guide the choice of sample size while curating empirical antibody data for ML applications.

\section{Conclusion and Open Challenges}
In this work, we methodically investigate whether pre-trained and fine-tuned language models are more effective at predicting antibody binding affinity properties and analyze data training sufficiency requirements. 
We demonstrate that using a powerful language model, the BERTTranformer, we can learn protein representations that consistently outperform other techniques.  Surprisingly, we discover that performance on the affinity binding task is best when the most diverse and general dataset (Pfam) is used in pre-training. This result supports the hypothesis that antibody structural and functional properties are informed by more than the antibody sequences alone; most probably also relying on additional features more broadly encountered in general protein sequences. 
Secondly, we find that training sets of sizes between 3200 and 6400 samples provide the highest performance gain in the affinity binding task. While performance gains continue as data size is increased, the cost/benefit of curating this additional data should be considered.

Our study has opened avenues for a few interesting research directions.
The initial hypothesis was that language models would be able to create context-dependent representations of antibody sequences and therefore would be better at predicting binding affinity. In this study, we have shown strong evidence of this hypothesis. Given these contextual representations, it is of great interest to understand what type of features are being learned by language models, do these features confirm known biological knowledge on antibody function, and finally, are the language representations generating novel features that highlight new insights about how the antibody sequence encodes structural and functional properties. 
More work is needed towards interpreting and biologically contextualizing learned features.

The ultimate goal of learning effective representations of antibody sequences is to support automation and acceleration of the drug design process. However, it is known that predicting high binding affinity does not necessary lead to antigen neutralization and furthermore, the desired manufacturing properties like stability are even harder to ensure based on binding affinity optimization alone. An exciting research direction addressing this challenge looks at multi-objective optimization, where properties in addition to binding are considered in the learning objective. 

\bibliographystyle{aaai22}
\bibliography{references}
\end{document}